\documentclass[prb,twocolumn,amsmath,amssymb,citeautoscript,showkeys,floatfix]{revtex4-1}

\usepackage{graphicx}
\usepackage{dcolumn}
\usepackage{bm}
\usepackage{lipsum}
\usepackage[most]{tcolorbox}
\usepackage[section]{placeins}
\usepackage{float}
\usepackage{mwe}
\usepackage{multirow}
\usepackage[flushleft]{threeparttable}
\usepackage{filecontents}

\begin{document}

\title{Trivial to nontrivial topology transition in rare-earth pnictides with epitaxial Strain} 		

\author{Shoaib Khalid}
\affiliation{Department of Physics and Astronomy, University of Delaware, Newark, DE 19716, USA}
\affiliation{Department of Material Science and Engineering, University of Delaware, Newark, DE 19716, USA}
\author{Anderson Janotti}
\email{janotti@udel.edu}
\affiliation{Department of Material Science and Engineering, University of Delaware, Newark, DE 19716, USA}


\begin{abstract}

The combination of magneto-transport and topological properties has brought great attention to rare-earth mono-pnictides semimetals. For some of them, like LaSb, it is unclear whether they show non-trivial topology or not based on density functional theory calculations and angular resolved photoemission spectroscopy measurements. Here, we use hybrid density functional theory to demonstrate that LaSb is in fact a trivial topological semimetal, in agreement with experiments, but on the verge of a transition to a topological phase. We show that under compressive epitaxial strain, the La $d$ band crosses the Sb $p$ band near the X$_3$ point in the Brillouin zone, stabilizing a topologically non-trivial phase, opening unique opportunities to probe the inter-relation between magneto-transport properties and the effects of band topology by examining epitaxially strained and unstrained thin films of the same material.  

\end{abstract}


\maketitle

\section{Introduction} \label{sec:intro}

Rare-earth monopnictides have attracted great attention for basic and applied sciences, 
displaying extreme magnetoresistance (XMR) and topologically non-trivial band structures\cite{chatterjee2019weak,yang2017extreme,tafti2016resistivity,sun2016large}, with applications in terahertz detectors\cite{bjarnason2004ErAs,salas2017growth}, solar cells \cite{gossard2007enhanced}, tunnel junctions\cite{zide2006increased}, and serving as epitaxial contacts to III-V semiconductors\cite{chatterjee2019weak,hanson2006ErAs,Zhang2019}. Most all rare-earth pnictides are semimetals that crystallize in the rocksalt structure, and due to the presence of partially filled $f$ states they are antiferromagnetic at low temperatures\cite{wakeham2016large,tsuchida1965magnetic,chattopadhyay1994high,petukhov1996electronic,mullen1974magnetic,li1997magnetic,petukhov1994electronic,khalid2020hybrid}, except for non-magnetic La, Y and Lu pnictides, where the $f$ shell is empty or completely full. 

\begin{figure}
\includegraphics[width=8.5 cm]{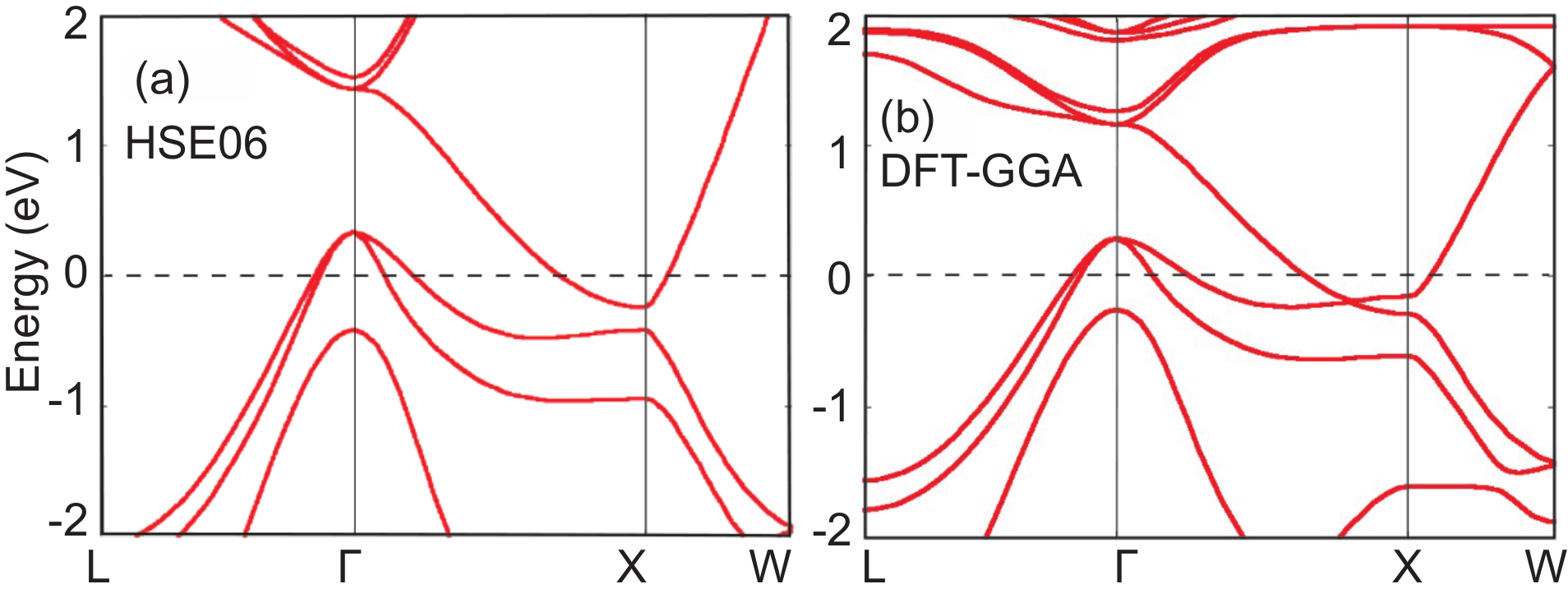}
\caption{(color online) Electronic band structure of LaSb using (a) HSE06 hybrid functional and (b) the DFT-GGA functional with spin-orbit coupling. The zero in the energy axis corresponds to the Fermi level.}
\label{fig1}
\end{figure}

In the past few years, many reports appeared on the topological properties of rare-earth pnictides \cite{nayak2017multiple,niu2016presence,lou2017evidence,khalid2019topological}. Insulators can be categorized into normal or trivial band insulators and non-trivial topological insulators in the presence of time reversal symmetry \cite{hasan2010colloquim,qi2011topological},
with the latter showing surface states that are spin-momentum locked and robust against any time reversal invariant local perturbation \cite{hasan2010colloquim,qi2011topological}. 
Similarly, semimetals can be classified in trivial and topological non-trivial, and the latter  are divided into Weyl, Dirac and nodal-line semimetals \cite{wan2011topological,weng2015weyl,liu2016evolution}. These topologically non-trivial semimetals are directly related to each other in a way that a Dirac semimetal transforms into a Weyl semimetal by splitting a Dirac point into two Weyl points provided that the time reversal symmetry or spatial inversion symmetry is lifted \cite{hirayama2018topological}. In analogy to the Dirac-cone surface states in the bulk band gap of topological insulators, the surface states of Weyl and Dirac semimetals are described by Fermi arcs \cite{liu2014discovery,xu2015observation,xu2015discovery,yang2015weyl}. The absence of bulk band gap in the non-trivial semimetals makes it more difficult to probe the Dirac-like cone surface states due to the overlap with the bulk states. 
These topological semimetals are characterized by the non-trivial $Z$\textsubscript{2} invariant using parity analysis provided that space inversion symmetry and time reversal symmetry are preserved and there exists a bulk band gap at each of the $k$ point in the Brillouin zone. 

Recent experiments on LaX (X=As, Sb and Bi) showed large magnetoresistance of up to 10\textsuperscript{5}\% with resistivity plateau at low temperatures \cite{tafti2016resistivity,yang2017extreme,sun2016large}, pointing to potential applications in sensor and spintronic devices \cite{daughton1999gmr,prinz1998magnetoelectronics,thomson1857xix,wu2016giant}.
Experiments and first-principles calculations indicated the presence of topological surface states in LaBi, while LaAs was clearly shown to behave as a trivial semimetal \cite{nayak2017multiple,niu2016presence,lou2017evidence,khalid2019topological}. Controversial results were reported for LaSb: first-principles calculations based on the density functional theory (DFT) within the generalized gradient approximation (GGA) predicted that LaSb is a topological semimetal with a crossing between the La $d$ and the Sb $p$ bands near the $X$ point \cite{nummy2018measurement}, while meta-GGA MBJ \cite{guo2016charge} and hybrid density functional calculations \cite{khalid2019topological} showed that such crossing does not occur, in agreement with ARPES measurements \cite{nummy2018measurement}. These results fuelled 
the debate of whether the observed magnetoresistance is due to the non-trivial topological  properties of the band structure of LaSb and LaBi, or due to complete compensation of the electron and hole pockets \cite{ali2014large,jiang2015signature,he2016distinct}.

In fact, LaX  (X=As, Sb and Bi) are compensated semimetals with equal electron and hole carrier concentrations \cite{sun2016large,zeng2016compensated,yang2017extreme}. LaBi is a non-trivial topological semimetal with three Dirac cones on the surface \cite{nayak2017multiple,lou2017evidence,nummy2018measurement}. LaAs is a trivial semimetal as shown in recent experimental studies \cite{yang2017extreme,nummy2018measurement}, but it can be made topologically non-trivial under hydrostatic pressure\cite{khalid2019topological}. Whether LaSb is a trivial or a topological semimetal and whether its band structure can be changed from trivial to topologically non-trivial by applying small perturbations are still matter of debate.  
The HSE hybrid functional gives an accurate description of the electronic structure of rare-earth pnictides, predicting carrier densities that are in very good agreement with experimental data \cite{chua1974simple,singha2017fermi,shirotani2003pressure,sun2016large,yang2017extreme,tafti2016resistivity}. It correctly describes the topologically trivial ground state of LaAs, and predicts the observed non-trivial topology of LaBi band structure \cite{nayak2017multiple,niu2016presence,lou2017evidence,khalid2019topological}. 

\begin{figure}
\includegraphics[width=8 cm]{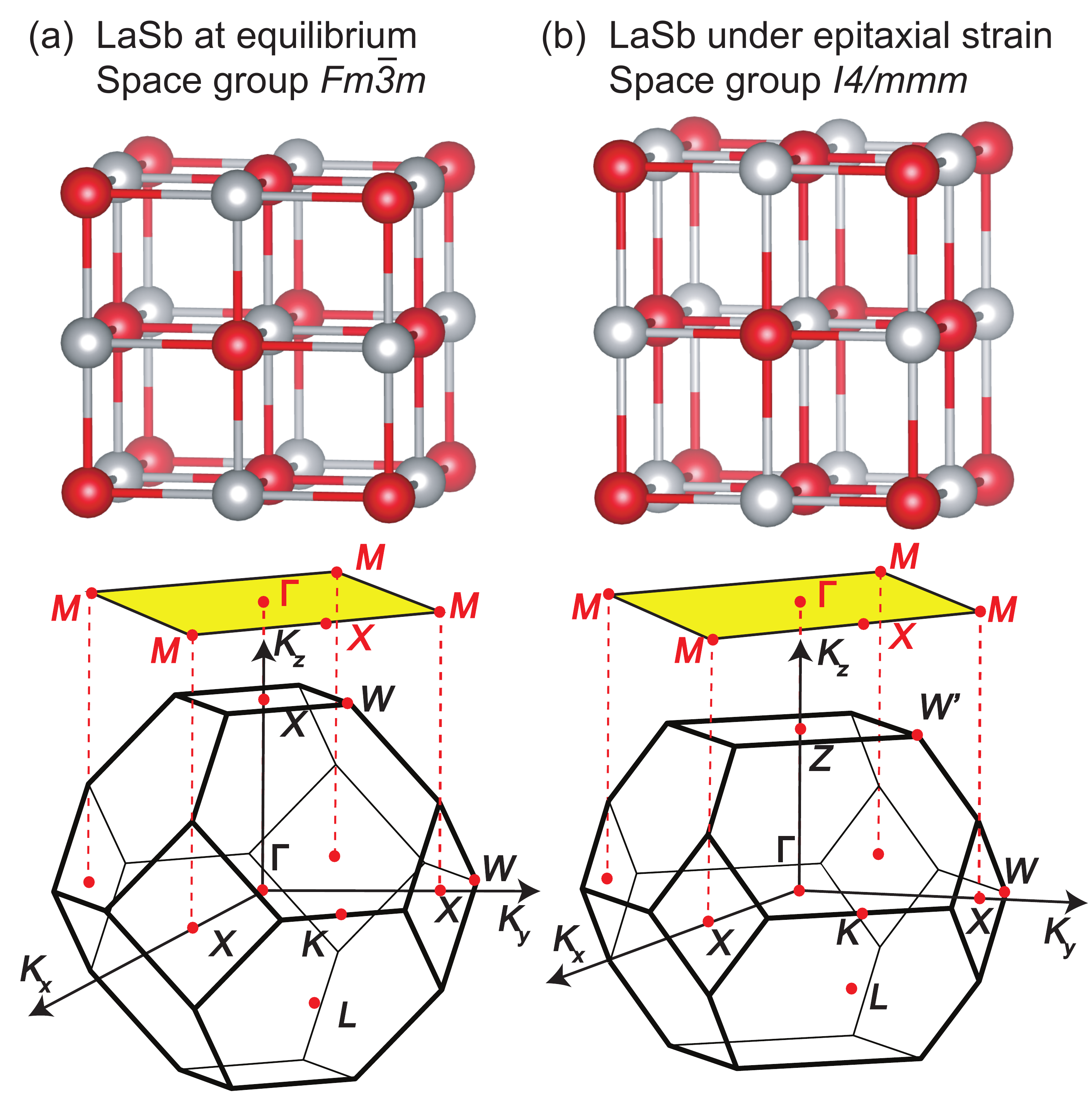}
\caption{(color online) Crystal structure and Brillouin zone of (a) LaSb at equilibrium and (b) under compressive epitaxial strain (the later was exaggerated for easy visualization). The shaded area in yellow shows the projection of the bulk Brillouin zone on the (001) surface Brillouin zone (SBZ), with symmetry points in the SBZ displayed in red.}
\label{fig2}
\end{figure}

Here we use HSE hybrid functional to describe the electronic structure of LaSb, and show that LaSb is indeed a semimetal with a trivial topological band structure. We then predict that LaSb can be turned into a topological semimetal under reasonably small epitaxial strain, reporting the bulk electronic structure of epitaxially strained LaSb and identifying the Dirac cone and spin texture of the surface states.

\section{Computational Approach} 
\label{sec:Com}

Our calculations are based on the density functional theory (DFT) \cite{hohenberg1964inhomogeneous,kohn1965self} with the projector augmented wave (PAW) method \cite{blochl1994projector,kresse1999ultrasoft} as implemented in the VASP code \cite{kresse1993ab,kresse1994ab}. For the exchange-correlation potential, generalized gradient approximation (GGA) of Perdew-Burke-Ernzerhof (PBE) \cite{perdew1996generalized} and the screened hybrid functional of Heyd, Scuseria, and Ernzerhof  (HSE06) \cite{heyd2003hybrid,HSE} are employed. In HSE06, the exchange-correlation functional is divided into long range and short range parts with the screening parameter set to $\omega$= 0.201 \AA$^{-1}$; a mixing of 25\% Fock-exchange \cite{perdew1996rationale} is used only in the short range part, while the long range part and the correlation are based on the PBE functional. We used PAW potentials for La with nine valence electrons,  5p\textsuperscript{6}6s\textsuperscript{2}5d\textsuperscript{1},  and five valence electrons for Sb, i.e., 5s\textsuperscript{2}5p\textsuperscript{3} configuration. A kinetic-energy cutoff of 300 eV is used for the plane wave basis set. 

For the calculation of epitaxial strain we first use an eight atom cubic unit cell to simulate the epitaxial strain (fixing $a$=$b$ and relaxing $c$), from which we extract a two atom primitive cell for band structure calculations, thus avoiding the effects of band folding. To sample the Brillouin zone we use 8$\times$8$\times$8 $\Gamma$-centered $k$-point mesh for the 2-atom cell calculations. Effects of spin-orbit coupling (SOC) are included only in the band structure calculations, not in the cell optimization. Owned to the inversion symmetry and time-reversal symmetry at equilibrium and under epitaxial strain, the $Z$\textsubscript{2} topological invariant can be calculated from the product of parities at the time-reversal invariant momentum (TRIM) points \cite{fu2007topological}. The wannier90 code \cite{mostofi2014wannier90} is used to obtain maximally localized Wannier functions (MLWF) and to parameterize a tight binding (TB) hamiltonian. The WannierTools code \cite{wu2018wannier} is used to obtain surface band structure and spin texture based on the TB hamiltonian interfaced with wannier90.

\section{Results and Discussion} 
\label{sec:result}

Like the other rare-earth pnictides, LaSb is stable in the rock-salt crystal structure with space group $Fm{\bar{3}}m$. The calculated equilibrium lattice parameter of LaSb is 6.540 {\AA} using DFT-GGA and 6.514 {\AA} using HSE06, in good agreement with experimental lattice parameter of 6.488 {\AA} \cite{chua1974simple}. The calculated electronic band structure of LaSb using DFT-GGA and HSE06 are shown in Fig.~\ref{fig1}, focusing on the region within $\pm$2 eV of the Fermi level. The partially occupied bands near $\Gamma$ (hole pockets) are derived mostly from Sb 5$p$ orbitals, whereas the partially occupied bands at the $X$ point are derived from La 5$d$ orbitals. With the inclusion of spin-orbit coupling the 3-fold degenerate bands at $\Gamma$ split into a doubly degenerate band and a nondegenerate band, whereas the La $d$ band at $X$ remains unchanged.

\begin{figure*}[t!]
\includegraphics[width=12 cm]{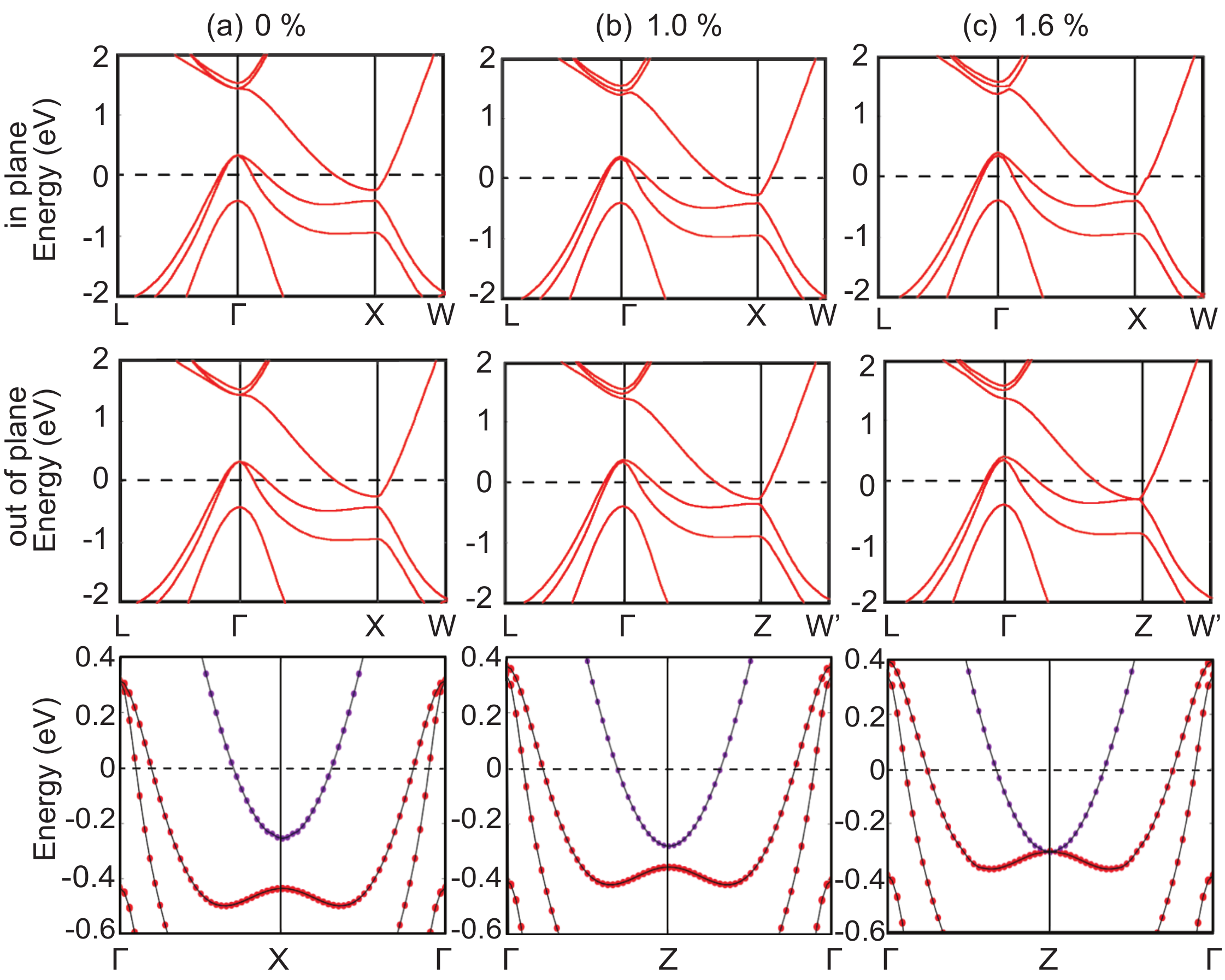}
\centering
\caption{(color online) Electronic band structures of LaSb with (a) 0\%  (b) 1\% and (c) 1.6\% applied compressive epitaxial strain. The top row shows the bands along the in-plane direction, the middle row shows the bands along an out-of-plane direction, whereas the bottom row shows the bands along the $\Gamma$-$X$-$\Gamma$ direction for the case of  0\% applied compressive epitaxial strain, and along the $\Gamma$-$Z$-$\Gamma$ direction for the cases of 1.0\% and 1.6\% applied compressive epitaxial strain all calculated using the HSE06 hybrid functional with spin-orbit coupling. The Fermi level is set to zero.}
\label{fig3}
\end{figure*}

The electronic structure of LaSb is still under debate as the calculations by Guo {\em et al.} \cite{guo2016charge} predict different results using DFT-GGA and meta-GGA (MBJ). The DFT-GGA functional predicts LaSb to be a topological semimetal with a band inversion at the $X$ point, whereas the meta-GGA predicts a topologically trivial behavior by decreasing the La $d$-Sb $p$ band overlap. HSE06 hybrid functional calculations by Guo {\em et al.} \cite{Guo2017theoretical} also predict a topologically trivial band structure. ARPES measurements by Nummy {\em et al.}\cite{nummy2018measurement} showed no sign of band inversion, whereas results by Niu {\em et al.} \cite{niu2016presence} were not conclusive regarding the non-trivial topological nature of LaSb.

Our DFT-GGA and HSE06 calculations show an overlap in energy between La $d$ band and Sb $p$ band thus confirming the semimetallic nature of LaSb in agreement with previous calculations \cite{guo2016charge,Guo2017theoretical} and experiments \cite{zeng2016compensated,nummy2018measurement,tafti2016resistivity}.
However, the HSE06 calculations reveal that the La $d$ band lies higher in energy than the Sb $p$ bands at each $k$ along the $\Gamma-X$ direction, while in the DFT-GGA there is a crossing between the La $d$ and Sb $p$ bands along the $\Gamma-X$ direction. 
We note that our HSE06 calculations correct the overestimated La $d$-Sb $p$ band overlap in DFT-GGA and also correctly describes the trivial nature of LaSb band structure in agreement with previous HSE06 calculations \cite{Guo2017theoretical} and experiments\cite{nummy2018measurement,zeng2016compensated}.
Note also that the order to the La $d$ band and the pnictogen $p$ band can be reversed if the spin-orbit coupling is stronger as in the case of LaBi \cite{khalid2019topological,nayak2017multiple,lou2017evidence}, which is a similar material with a topologically non-trivial band structure. 

The electronic structure of a material can be changed without adding any extraneous chemical species or doping. Recent studies have shown that applying hydrostatic pressure leads to an inversion of the La $d$ and the pnictogen $p$ bands at the $X$ point in LaAs\cite{khalid2019topological} and LaSb\cite{Guo2017theoretical}, turning them into topologically non-trivial materials. Epitaxial strain, such as in coherently strained films grown or deposited on lattice mismatched substrates, can also be employed to alter the electronic structure of materials. It reduces the space group symmetry from $Fm{\bar{3}}m$ to $I4/mmm$, yet keeping the inversion symmetry. Here we show that compressive epitaxial strain can be used as an effective tool to change the topological properties of the LaSb band structure. Epitaxial strain has been demonstrated in the growth of rare-earth pnicitides on  III-V semiconductors using molecular beam epitaxy \cite{chatterjee2019weak,Shouvik2020}. Compressive epitaxial strain of up to 3\% has already been realized in III-V semiconductors\cite{Richardson2016}, and we expect that similar strains can be achieved in the rocksalt rare-earth pnictides. 
Note that effects of charge transfer at the interface between the substrate and the LaSb thin film may affect the carrier compensation; the amount of charge will depend on the band alignment, the interface termination, and the doping type and level in the semiconductor.

In Fig.~\ref{fig2} we show how the Brillouin zone of LaSb changes under epitaxial strain. Under equilibrium, the Brillouin zone has the shape of a truncated octahedron, which is common for face-centered cubic or rocksalt crystal structures, with $\Gamma$ at the center and three independent $X$ points at the center of the square faces. The shape of the Brillouin zone changes when we apply compressive epitaxial strain as the three independent $X$ points split into two in-plane $X$ points and one out-of-plane $Z$ point. It is important to note that the spacial inversion symmetry is preserved in LaSb under epitaxial strain, and it is reflected in the inversion symmetry of the distorted Brillouin zone shown in Fig.~\ref{fig2}(b).

The calculated electronic band structure of LaSb under compressive epitaxial strain is shown in Fig.~\ref{fig3}, along the $\Gamma-X-\Gamma$ and $\Gamma-Z-\Gamma$ directions.
Using HSE06 functional we find that under equilibrium  the 3 independent $X$ points are equivalent, and the band structure along the in-plane and out-of-plane  $\Gamma-X$ directions are the same. However, under epitaxial strain the 3 independent $X$ points are split into two in-plane $X$ and one out-of-plane $Z$ point.
In the case of compressive epitaxial strain of 1\%, the La $d$ band gets closer to the Sb $p$ bands as the overall volume of the crystal is decreased. However, the La $d$ and Sb $p$ bands along the out-of-plane $\Gamma-Z$ direction become even closer compared to those along the in-plane $\Gamma-X$ direction, but still there is no crossing between the La $d$ and Sb $p$ bands.
At 1.6\% compressive epitaxial strain we observe a band inversion along the out-of-plane $\Gamma-Z$ direction, whereas the La $d$ and Sb $p$ bands along the $\Gamma-X$ still avoid crossing.

Note that the crossing between the La $d$ and Sb $p$ bands only occur in the case of compressive epitaxial strain, not for tensile strain. For compressive strain, the La $d$ band and Sb $p$ bands approach each other along both the $\Gamma-Z$ and $\Gamma-X$ directions because the overall volume/cell of the crystal decreases. The out-of-plane $c$  lattice parameter changes according to the calculated Poisson's ratio, $\nu=-\frac{\varepsilon_{zz}}{\varepsilon_{xx}}=0.26$, where $\varepsilon_{xx}=\varepsilon_{yy}$ is the change of the lattice parameters along the in-plane $a$ and $b$ directions, and ${\varepsilon_{zz}}$ is the change of the lattice parameter along the out-of-plane $c$ direction under epitaxial strain. For tensile epitaxial strain, the La $d$ and Sb $d$ bands are further separated from each other due to the overall increase in the volume/cell of the crystal.

To shed light on the bulk band inversion and its relationship with the non-trivial topology in LaSb under epitaxial strain, we calculated the $Z$\textsubscript{2} invariant,
which is given by the product of parities of all the occupied bands at the TRIM points\cite{fu2007topological} through the following relation:
\begin{equation}
(-1)\textsuperscript{$v$\textsubscript{0}}= \prod_{m=1}^{8} \delta_{m}
\label{eq1}
\end{equation}
where $\nu_0$ indicates the topological nature of the material and $\delta_{m}$ is the parity product at $m$-th TRIM point for all the occupied bands. For the La$X$ ($X$=As,Sb,Bi) compounds, this $Z$\textsubscript{2} invariant actually depends upon the product of parities of the occupied bands at the $X$ points, which in the case of epitaxial strain will be $X$ and $Z$ points.

\begin{table}
\setlength{\tabcolsep}{7pt}
\caption{Parities of all the occupied bands at TRIM points in the first Brillouin zone of LaSb under 0\% applied compressive epitaxial strain.
}
\centering
\begin{tabular}{l c c c c c c c c c}
\hline\hline
No.  &$\Gamma$  & $L$ & $L$  & $L$ & $L$ & $X$ & $X$ & $X$ & Total
        \\  
\hline 
1 & - & - & - & - & - & - & - & - & + \\[0.5ex]
3 & - & - & - & - & - & - & - & - & +  \\[0.5ex]
5 & - & - & - & - & - & - & - & - & +  \\[0.5ex]
7 & + & - & - & - & - & + & + & + & +  \\[0.5ex]
9 & - & + & + & + & + & - & - & - & +  \\[0.5ex]
11 & - & + & + & + & + & - & - & - & +  \\[0.5ex]
13 & - & + & + & + & + & - & - & - & +  \\[0.5ex]
Total & + & + & + & + & + & + & + & + & {\large\textbf{+}} \\[0.5ex]
\hline\hline 
\end{tabular}
\label{tab1} 
\end{table}
\begin{table}
\setlength{\tabcolsep}{7pt}
\caption{Parities of all the occupied bands at TRIM points in the first Brillouin zone of LaSb under 1.6\% applied compressive epitaxial strain).
}
\centering
\begin{tabular}{l c c c c c c c c c}
\hline\hline
No.  &$\Gamma$  & $L$ & $L$  & $L$ & $L$ & $X$ & $X$ & $Z$ & Total
        \\  
\hline 
1 & - & - & - & - & - & - & - & - & + \\[0.5ex]
3 & - & - & - & - & - & - & - & - & +  \\[0.5ex]
5 & - & - & - & - & - & - & - & - & +  \\[0.5ex]
7 & + & - & - & - & - & + & + & + & +  \\[0.5ex]
9 & - & + & + & + & + & - & - & - & +  \\[0.5ex]
11 & - & + & + & + & + & - & - & - & +  \\[0.5ex]
13 & - & + & + & + & + & - & - & + & -  \\[0.5ex]
Total & + & + & + & + & + & + & + & - & {\large\textbf{-}}  \\[0.5ex]
\hline\hline 
\end{tabular}
\label{tab2} 
\end{table}

In LaSb at equilibrium, the valence band is mostly composed of Sb $p$ orbitals, whereas the conduction band has most of the contributions from La $d$ orbitals. Parity of  La $d$ band is X$^{\mathrm{+}}_{\mathrm{7}}$ which is even, while the parity of Sb $p$ band is X$^{\mathrm{-}}_{\mathrm{7}}$ which is odd at two inequivalent $X$ point and one $Z$ point as shown in Table ~\ref{tab1}. Under compressive epitaxial strain of $\sim$1.6\%, the parity at the $Z$ point is interchanged, whereas the parity at the two in-plane $X$ points remains unchanged as shown in Table ~\ref{tab2}. Due to the change in parity at the $Z$ point under applied compressive epitaxial strain the $Z$\textsubscript{2} invariant switches from 0 to 1 which is a sign of a non-trivial topological semimetal.

\begin{figure*}
\includegraphics[width= 15 cm]{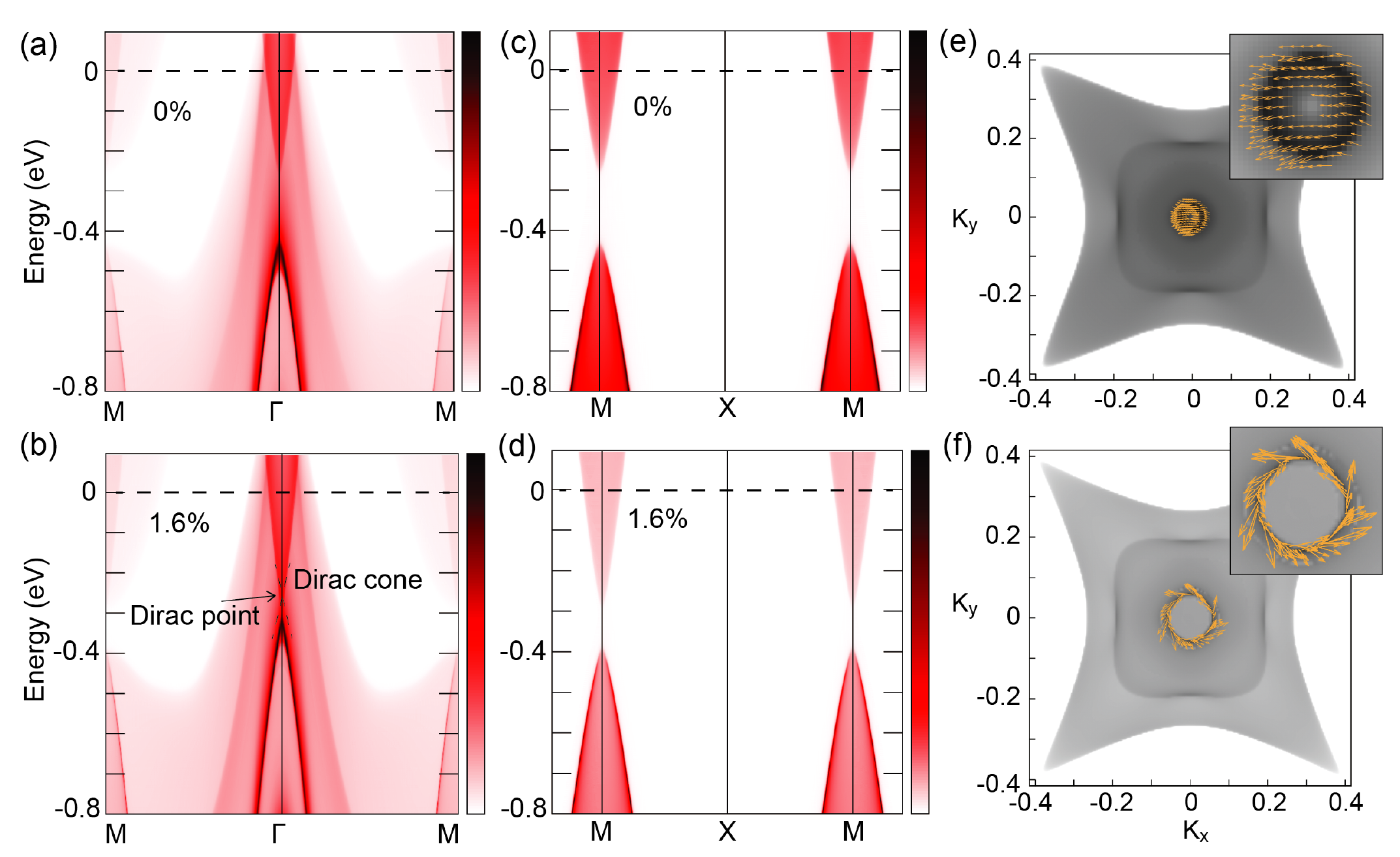}
\caption{(color online) Electronic band structure of (a) LaSb at equilibrium along $M$-$\Gamma$-$M$ and (c) $M$-$X$-$M$ directions and  (b) LaSb under 1.6\% compressive epitaxial strain along $M$-$\Gamma$-$M$ and (d) $M$-$X$-$M$ directions in the (001) surface Brillouin zone (SBZ). The bar on the right shows the intensity of surface states. The Fermi level is set to zero. The Dirac cone and Dirac point are indicated in (b). The spin textures of the bands crossing at 0.5 eV below the Fermi level of LaSb under equilibrium and LaSb under 1.6\% compressive epitaxial strain are shown in (e) and (f), respectively.}
\label{fig5}
\end{figure*}

To further characterize the topologically non-trivial nature of LaSb under epitaxial strain we calculated the (001) surface band structure. From previous work \cite{nayak2017multiple} we learned that LaBi is a topologically non-trivial material under equilibrium. LaBi has bulk band inversion at three inequivalent $X$ points, so that the surface states show three Dirac cones related to the bulk band inversions that are projected onto the surface Brillouin zone. Two in-plane $X$ points are projected onto $M$ points in the surface Brillouin zone (SBZ), and the third, out-of-plane, $X$ ($Z$) point is projected onto $\Gamma$. 

Since the bulk band inversion in LaSb under the effect of compressive epitaxial strain occurs only at the $Z$ point (i.e., the out-of-plane $X$ point), we expect only one Dirac cone to appear at the surface, located at the $\Gamma$ point. In Fig.~\ref{fig5}(a) and (b) we show the LaSb(001) surface electronic band structure along $M$-$\Gamma$-$M$ direction. We see that a Dirac cone is missing at the $\Gamma$ point when LaSb is at equilibrium, indicating of the trivial behavior in agreement with previous results \cite{guo2016charge,Guo2017theoretical,nummy2018measurement,zeng2016compensated}. However, a gapless Dirac cone appears at the $\Gamma$ point in LaSb under compressive epitaxial strain of 1.6\%, in agreement with parity analysis. In Fig.~\ref{fig5}(c) and (d) we also show the surface band structure along the $M$-$X$-$M$ direction and, as expected from the discussion above, there is no sign of surface Dirac cones even under compressive epitaxial strain. 

Another feature of non-trivial band structures is the presence of helical spin texture. To demonstrate that, we determined the spin texture for the surface band structures of LaSb at equilibrium and under 1.6\% compressive epitaxial strain, the results of which are shown in Fig.~\ref{fig5}(e) and (f), respectively. The spin textures are calculated at an energy cut of 0.5 eV below the Fermi level. In Fig.~\ref{fig5}(e), we see that the helical spin texture is missing, with all spins aligned in one direction, whereas in Fig.~\ref{fig5}(f) we observe a helical spin texture, further corroborating our analysis above. These calculations clearly demonstrate that under reasonably small compressive epitaxial strain, of 1.6\%, LaSb becomes a topological semimetal. The ability to coherently grow and characterize thin films of LaSb on lattice matched and lattice mismatched substrates that offer such relatively small strains could shed light on the role of carrier compensation and non-trivial topology in the observed extreme magnetoresistance effects.
 
\section{Summary} 
\label{sec:sum}

Using first-principles calculations we studied the electronic structure of LaSb using DFT-GGA and the screened hybrid functional HSE06. We verify that HSE06 rectifies the overestimated band overlap between conduction and valence bands in DFT-GGA. Using HSE06 we showed that LaSb is a topologically trivial semimetal under equilibrium, in agreement with the experiments. We also show that compressive epitaxial strain can be used to turn LaSb films into topological semimetal, creating unique opportunity to probe the inter-relationship between the occurrence of non-trivial topological properties, compensation of electrons and holes, and extreme magnetoresistance in rare-earth pnictides. 

\section*{Acknowledgements}
We acknowledge fruitful discussions with C.J. Palmstrøm and S. Chatterjee. 
This work was supported by the U.S. Department of Energy under Award
No.~de-sc0014388, and it used resources of the National Energy Research Scientific Computing Center (NERSC), a U.S. Department of Energy Office of Science User Facility operated under Contract No. DE-AC02-05CH11231.

%

\end{document}